\documentclass[11pt]{article}

\usepackage{amsfonts}
\usepackage{amsmath}
\usepackage{latexsym}
\usepackage{amssymb}
\usepackage[mathcal]{euscript}

\begin{document}

\begin{center}
\LARGE{\textsc{{Kaigorodov spaces \\ and their Penrose limits}}}
\end{center}
\medskip
\begin{center}
Christophe Patricot \footnote{email: \texttt{C.E.Patricot@damtp.cam.ac.uk}}\\
\end{center}
\begin{center}
\emph{ DAMTP,\\ University of Cambridge,\\ Centre for Mathematical Sciences,\\
Wilberforce Road,\\ Cambridge CB3 0WA,\\ U.K.}
\end{center}



\begin{abstract}
Kaigorodov spaces arise, after spherical compactification, as near horizon limits of M2, M5, and D3-branes with a particular pp-wave propagating in a world volume
direction. We show that the
uncompactified near horizon configurations $K\times S$ are solutions
of $D=11$ or $D=10$ IIB supergravity which correspond to
perturbed versions of their $AdS \times S$ analogues. We derive the
Penrose-G\"uven limits of the Kaigorodov space and the total spaces
and analyse their symmetries. An In\"on\"u-Wigner contraction of the Lie algebra is shown to
occur, although there is a symmetry enhancement. We compare the results to the maximally supersymmetric CW spaces found as
limits of $AdS\times S$ spacetimes: the initial gravitational perturbation on the
brane and its near horizon geometry remains after taking
non-trivial Penrose limits, but seems to decouple. One particular
limit  yields a time-dependent
homogeneous plane-wave background whose string theory is solvable,
while in the other cases we find inhomogeneous backgrounds.  
\end{abstract}

\newpage


\section{Introduction}
\label{intro kaigo}
Penrose showed \cite{Penrose plane-wave} that any spacetime, in the
neighbourhood of a null geodesic containing no conjugate points, has a
plane wave limit spacetime. This limit is essentially local, and may
be thought of physically as the spacetime seen by an observer approaching the
speed of light at a given point,  along a particular null
geodesic, and recalibrating his clock to its affine parameter. It can thus be understood as a null
Lorentz boost together with a (singular) uniform rescaling of the
coordinates which leaves the affine parameter along the chosen null
geodesic invariant. G\" uven \cite{Gueven} extended the concept to
supergravity theories: given a solution 
of the supergravity equations of motion, there exists a limit of this
solution which also satisfies these
equations and has a plane-wave spacetime. \\
It has long been known that string theory on gravitational wave backgrounds    
is exact and potentially solvable \cite{Horowitz}, mainly because all
of the curvature invariants of these spaces vanish
\cite{Gary}. The recently discovered \cite{BFHP} (BFHP) maximally
supersymmetric plane-wave solution of IIB string theory
was shown to be exactly solvable \cite{Metsaev}. The fact that this
maximally supersymmetric plane-wave, together with its 11-dimensional analogue
\cite{Kowalski} (KG),  arise
as Penrose limits of $AdS\times S$ spacetimes \cite{Blau}, suggests
one could probe into the  string
or M-theories of the latter. This idea was vindicated by the BMN proposal
\cite{BMN}, or plane-wave/CFT correspondence. \\
One should not forget however that $AdS\times S$
spacetimes, though maximally supersymmetric solutions of supergravity
or IIB string theory by themselves, also essentially arise in these theories as
near-horizon limits of extremal branes. These branes also are
fundamental dynamical objects which maybe in essence should be thought
of as canditates for ``world-volumes''. The AdS/CFT
correspondance together with the BMN correspondance certainly give an insight into gravity theories on the
extremal branes whose near horizon geometry is of $AdS\times S$ type,
but what if the geometry of these branes were perturbed ? \\
In \cite{Cvetic} and \cite{Chamblin}, the
near-horizon geometries of non-dilatonic extremal branes with a pp-wave propagating in the
world-volume were analysed, and it was found that in the most simple
case one gets a product of a sphere and a homogeneous space of constant negative
cosmological constant which generalizes the 4 dimensional Kaigorodov
space \cite{Kaigorodov}. We will show in fact that the near horizon
configurations of these particular M2, M5 and D3-pp-branes are also
solutions of D=11 supergravity or D=10 IIB supergravity, and are
analogous to the $AdS \times S$ spaces. Therefore it seems interesting to find the
Penrose limits of these spaces, not for the mere sake of
geometry, but rather to find backgrounds to study string theory which
naturally arise from perturbed branes.\\ 
Indeed, string theory was explicitly 
solved in the light-cone gauge not only in the time-independent
Cahen-Wallach  spaces, but also recently in some time dependent
plane-wave backgrounds (\cite{Fuji}\cite{Russo}, and
earlier references therein). The
quantization of strings and particles gives rise to time-dependent
harmonic oscillators, and whenever the plane-wave spaces are homogeneous,  it
seems the equations of motion can be solved explicitly (for the
particle case see \cite{Loughlin}). It turns out
that one of the plane-wave spaces we derive is very similar to that
in \cite{Russo}, and on these grounds should admit an explicitly
solvable string theory,  though one of the oscillators has negative
mass. The article is organized as follows.\\

In section \ref{Kaigo in M-theory} we first briefly review how
Kaigorodov spaces arise as near horizon limits of M2, D3, and M5 pp-branes \cite{Cvetic}.  We show
that the near-horizon geometries of type $K_{n+3} \times S^d$ together with their
fluxes, satisfy the equations of supergravity or IIB theory in the
respective cases. In fact, these configurations can be thought of as
perturbed  $AdS\times S$ spaces. We then briefly describe the geometry of the Kaigorodov
 space $K$. This non-conformally flat homogeneous Einstein spacetime can be
interpreted \cite{Podolsky} as an $AdS$ space with a propagating
gravitational wave. Although it is non-static and has a
pp-singularity, we show that it is stably causal in the sense of
\cite{Hawking}. In
fact, the gravitational wave will be reminiscent in the non-trivial
Penrose limit spaces we find. In section \ref{P limits of K} we derive the  Penrose limit
 plane-wave spaces of the Kaigorodov space alone,  and analyse the
 symmetries of a particular limit space which is homogeneous. An In\" on\" u-Wigner contraction of the group of bosonic
 symmetries is shown to occur although there is an enhancement of 1 symmetry in the limit. We then derive in section
 \ref{general penrose limits} the Penrose limits of the uncompactified
 $K_{n+3} \times S^d$ spacetimes and their associated field-strengths, by
 considering both the null geodesics which wind around the
 sphere and those which do not. As opposed to the $AdS \times S$ case,
we get non-trivial 
 limits even in the non-winding case. These provide homogeneous
plane-wave backgrounds with
solvable string theories which have vanishing field strength and
constant dilaton. The non-winding case yields
non-homogeneous plane-waves whose matrix $A_{ij}$ in Brinkman
coordinates is not diagonal.      

\section{Kaigorodov spaces in M/String theory}
\label{Kaigo in M-theory}

\subsection{Near-horizon limits of perturbed branes}
We first review in detail the near horizon geometry of an M2-brane
with a gravitational wave propagating in one of its world volume
directions, and then give the general expressions of the metrics and
derive the field strengths of the near horizon geometries in  the 3
cases of the M2,
D3, and M5 pp-branes. The pp-wave or brane-wave solutions actually
arise as intersections of branes \cite{Tseytlin} \cite{Klebanov}.\\ 
Following \cite{Cvetic}, a  $D=11$ supergravity solution describing  a
non-dilatonic extremal M2-brane with a pp-wave is given by
\begin{eqnarray*}
ds^{2}_{11} &=& H^{-2/3}(-K^{-1}dt^{2}+K(dx_{1}+(K^{-1}-{1})dt)^{2}+dx^{2}_{2})+H^{1/3}(dr^{2}+r^{2}d\Omega^{2}_{7}),\\
F_4 &=& dt\wedge dx_{1}\wedge dx_{2}\wedge dH^{-1},\\
H &=& 1+\frac{Q_{1}}{r^{6}}, \qquad K=1+\frac{Q_{2}}{r^{6}},
\end{eqnarray*}
where $r$ is the distance to the brane in the bulk. The affine change of coordinates $t'+x'/2=t ,
t'+3x'/2=x_{1}$ yields a simpler expression, similar to the formalism
used in \cite{Chamblin}:
\begin{eqnarray}
\label{eq M2}
ds^{2}_{11} &=&
H^{-2/3}(2dt'dx'+(K+1){dx'}^{2}+dx^2_2)+H^{1/3}(dr^{2}+r^{2}d\Omega^{2}_{7}),\\
F_4 &=& dt'\wedge dx'\wedge dx_{2}\wedge dH^{-1},\nonumber
\end{eqnarray}
and one can change $K$ into $K-1$ by further coordinate transformation. This shows that
from the brane point of view, the gravitational wave propagates
along
the 
$t'$ null direction and is uniformly distributed along the
world-volume directions, but although this solution is often denoted as a pp-brane, the
metric (\ref{eq M2}) does not possess a covariantly constant null Killing
vector: it merely arises as the intersection of a membrane with an 11
dimensional pp-wave. Note that the wave is not localized on the brane
since $K(r)$ depends on $r$, but spacetime is asymptotically flat away
from the brane. The near horizon limit $r \longrightarrow 0$ together
with rescalings of the coordinates by powers of $Q_1$ and $Q_2$, and the change of variable  $\rho =\ln
r$, induce the following metric and field strength:
\begin{eqnarray}
\label{eq KcrossS}
ds^2_{11}&\sim &
Q^{1/3}_1\Big(e^{-2\rho}{dx''}^2+e^{4\rho}(2dt''dx''+{dx''_2}^2)
+d\rho^2+d\Omega^2_7\Big) \\
F_4 &\sim &6Q_1^{1/2}e^{6\rho}dt''\wedge dx''\wedge dx''_2\wedge d\rho
\end{eqnarray}
This is the  metric of the product space of a 4 dimensional Kaigorodov
space \cite{Kaigorodov} of negative cosmological constant $\Lambda = -12Q^{-1/3}_1$ and
a 7-sphere of radius $R^2~=~Q^{1/3}_1$. It will be denoted $K_4 \times S^7$.\\ 
The cases of the near horizon geometries of the M-5 supergravity brane
and the D-3 type IIB brane are obtained in a similar way. They yield $K_{n+3}
\times S^d$ spaces, with $K_{n+3}$ the $n+3$-dimensional
generalisation of $K_4$, and $S^d$ the $d$-dimensional sphere. The
(negative) cosmological constant $\Lambda $ of $K_{n+3}$ and the radius
of the sphere $R_{S^d}$ depend on the
charges of the initial branes. It turns out these can be expressed quite simply
combining results of \cite{Cvetic} and \cite{Chamblin}. 

\subsection{$K_{n+3}\times S^d$ spaces as solutions of supergravity}
In $n+3+d=11$,
$10$ or $11$ dimensions, for $n=1$, $2$, $4$ 
respectively (M2, D3, M5 pp-brane), the near horizon geometry is a
(topological) product of the Kaigorodov space $K_{n+3}$ of cosmological
constant $\Lambda= -4(n+2)L^2$ and a $d$-sphere $S^d$ of radius
$R=1/(Ln)$, with $L$ depending on the charges of the brane. If we call $Q$ the charge of the extremal brane, in the sense that
its harmonic function is $H(r)=1+Q/r^{(d-1)}$, then the Kaigorodov
parameter $L$ of the horizon geometry is $L=(1/n)Q^{-\frac{1}{d-1}}$,
with $d$ the dimension of the $S^d$. After rescaling the
colatitude coordinate $\psi$ of $S^d$ by a factor of $1/(Ln)$, the
3 metrics read:  
\begin{equation}
\label{totalspace}
ds^2_{n+3+d}=e^{-2Ln\rho}dx^2+e^{4L\rho}(2dxdt+dy^idy^i)+d\rho^2+d\psi^2+(Ln)^{-2}\sin^2\!(Ln\psi)
d\Omega^2_{d-1}
\end{equation}
Here $d\Omega^2_{d-1}$ is the surface element of a $(d-1)$-sphere of
unit radius, and $i\in\{1, \ldots, n \}$.  Neither \cite{Cvetic} nor
\cite{Chamblin} give the expressions of the fluxes in the limit. In
our coordinate system, restoring $n=1$, $2$, $4$ for the M2, D3,
M5-pp-branes respectively, the
field strengths are :
\begin{eqnarray}
F_4 &=& 6Le^{6L\rho}d\rho\wedge dx\wedge dt\wedge dy^1 \qquad
(K_4\times S^7) \nonumber \\
F_5 &=& 8Le^{8L\rho}d\rho\wedge dx\wedge dt\wedge d^2y^i +\ast [
8Le^{8L\rho}d\rho\wedge dx\wedge dt\wedge d^2y^i] \quad (K_5
\times S^5) \nonumber \\
\ast [F_4] &=& 12Le^{12L\rho}d\rho\wedge dx\wedge dt\wedge d^4y^i   
\quad (K_7 \times S^4) \nonumber \\
\label{fluxes} 
\end{eqnarray}
The field equations of supergravity \cite{Julia} with the fermionic
fields set to zero,
in the conventions used in \cite{O'Farrill}, read:
\begin{align}
R_{MN} &=\frac{1}{12}(F_{MPQR}{F_N}^{PQR}
-\frac{1}{12}g_{MN}F_{PQRS}F^{PQRS}) \label{Einstein} \\
\textrm{d}F &=0 \label{closed} \\
\textrm{d}(\ast F)&=\frac{1}{2} F \wedge F \label{co-closed}
\end{align}
Calling $g(K_{n+3})$ the determinant of the metric of the Kaigorodov space $K_{n+3}$, we
see that $\sqrt{-g(K_{n+3})}= e^{2L(n+2)\rho}$. Thus the flux $F_4$ is
proportional to the  volume form on $K_4$ arising from the metric,
hence is closed (\ref{closed}) and co-closed. As $F_4 \wedge F_4=0$,
$F_4$ satisfies (\ref{co-closed}). The Einstein equation
(\ref{Einstein}) follows  since $K_4$ and $S^7$ are Einstein
spaces. In detail, with $\mu,\nu$ labelling the coordinates of $K_4$ and
$a,b$ those of $S^7$, the right-hand-side of
(\ref{Einstein}) expands to:
\begin{align*}
R_{\mu \nu}&= -18L^2g_{\mu\nu} +6L^2g_{\mu \nu} \\
R_{ab}&=6L^2 g_{ab} 
\end{align*}
For Kaigorodov spaces $R_{\mu \nu}= -4(n+2)L^2 g_{\mu \nu}$ (see next
section) and for
$d$-dimensional spheres of radius $R$,
$R_{ab}=(d-1)R^{-2}g_{ab}$, but here $R=(Ln)^{-1}$. Hence (\ref{Einstein}) is satisfied, and 
$(K_4 \times S^7, F_4)$ is a solution of supergravity. \\
Similarly,  since $\ast[F_4]$ is proportional
to the
volume form of $K_7$,  $\widetilde{F_4}\equiv -\ast( \ast [F_4])$ is
closed and co-closed, and proportional to the volume form of
$S^4$. In fact $\widetilde{F_4}= +12L \textrm{Vol}_{S^4}$, thus
$R_{\mu \nu}= -24L^2 g_{\mu \nu}$ (on $K_7$) and
$R_{ab}=48L^2g_{ab}=(4-1)(4L)^2$ (on $S^4$). Hence $(K_7 \times S^4, \widetilde{F_4})$ is a solution of
supergravity. $F_5$ is proportional to   the self dualized volume form of
$K_5$ or $S^5$,  and the same computations show that $(K_5 \times
S^5, F_5)$ is a  solution of the field equations \cite{Schwarz} of chiral $N=2$ $D=10$ supergravity. These solutions are completetly analogous to the 3 $AdS_{n+3} \times
S^d$ configurations, in the same way as the M2, M5 and D3-pp-branes are
analogous to their ``flat'' versions. We now review some important
geometric features of Kaigorodov spaces.

\subsection{Kaigorodov spaces versus $AdS$ spaces}
Some general properties of Kaigorodov spaces can be found in
\cite{Cvetic} and \cite{Podolsky}. In
$n+3$ spacetime dimensions, letting $L=\frac{1}{2}\sqrt{-\Lambda /(n+2)}$, their metric
reads:
\begin{equation}
\label{kaigo}
ds^2_{n+3} = e^{-2nL\rho}dx^2+e^{4L\rho}(2dxdt+\sum_{i=1}^n{dy^i}^2)+d\rho^2
\end{equation}
These spaces are solutions of Einstein pure gravity with cosmological
constant $\Lambda$; they admit $\frac{1}{2}n(n+3)+3$ Killing vectors and are
homogeneous spaces. They are not static, and are $1/4$
supersymmetric. We show in section \ref{plk} that they are stably
causal in the sense of Hawking \cite{Hawking}. \\
The change of coordinates
\begin{displaymath}
z=e^{-\rho /R}, \quad x=Rx^+,\quad t=Rx^-,\quad y^i=Rx^i
\end{displaymath}
where $R=1/(2L)$, takes the metric to a horospherical-type form ($z>0$ only)
\begin{equation}
\label{horo}ds^2_{n+3}=\frac{R^2}{z^2}\big(2dx^+dx^- +z^{n+2}(dx^+)^2
+dx^idx^i +dz^2\big)
\end{equation}
In this appendix, starting from this expression of the metric, we find
an isometric embedding of the Kaigorodov space $K_{n+3}$ into a space
of signature $(2,n+3)$. In this coordinate system, the metric of the
uncompactified $K_{n+3}\times S^d$ spaces (\ref{totalspace}), with the
radii of the spheres $1/(Ln)$, read:
\begin{displaymath}
ds^2_{n+3+d}=\frac{R^2}{z^2}\big(2dx^+dx^- +z^{n+2}(dx^+)^2
+dx^idx^i +dz^2\big) + \Big(\frac{2}{n}\Big)^2R^2(d\widetilde{\psi}^2
+(\sin\widetilde{\psi})^2 d\Omega_{d-1}^2)
\end{displaymath}
Podolsk\'y \cite{Podolsky} argued  that the Siklos spaces (a general
family of spacetimes containing the
Kaigorodov space) can be viewed as an $AdS$ space-time with a
propagating gravitational wave, whose spatial direction rotates at a
constant velocity in orthonormal frames parellely transported along
timelike geodesics. He also showed that these spaces have a
pp-singularity at $z=+\infty$: the geodesic deviation equation becomes
singular, but the square of the Riemann tensor remains finite. This
occurs at $r=0$ in (\ref{eq M2}), which corresponds to the brane horizon
for us. We shall show in section \ref{geometry of the pp-wave} that
this pp-singularity, which represents the divergence of tidal forces as one approaches
the brane, can remain after taking the Penrose limit. \\

Clearly (\ref{horo}) can be interpreted as a wave propagating on the
horospheres of $AdS_{n+3}$, but $\partial_{x^-}$ is not covariantly
constant. However this form of the metric suggests the rescalings:
\begin{equation}
\label{infinite boost}
(x^-,\;x^+,\; x^i,\; z)\longmapsto (x^-, \; \Omega^2 x^+,\; \Omega x^i,
\; \Omega z )  
\end{equation}
whereupon taking the singular limit $\Omega \to 0$ the metric becomes that
of $AdS_{n+3}$ in horospherical coordinates
\begin{displaymath}
ds_{n+3}^2=\frac{R^2}{z^2}\big(2dx^+dx^- +dx^idx^i +dz^2\big)
\end{displaymath}
Although $\partial_{x^-}$ is null, Killing and hence geodetic, this is
not a Penrose limit, since the metric is not rescaled by $\Omega^{-2}$
to yield conservation of the affine parameter along $\partial_{x^-}$. The
corresponding Penrose limit, as explained in the next section, yields
flat space. Thus (\ref{infinite boost}) can be interpreted as an
infinite unrescaled boost of the Kaigorodov spacetime, which yields
$AdS$. The dynamical interpretation of this boost is unclear though. (One also gets the $AdS$ metric by rescaling all the
coordinates by $\Omega$ and letting $\Omega \to 0$). In
\cite{Chamblin}, it is shown that the boundary CFT energy-momentum
tensor of the Kaigorodov space is a constant null momentum density,
therefore it is the Kaigorodov space which should be thought of as an
infinitely boosted $AdS$ spacetime, and not the reverse. It will be 
more relevant in our analysis to consider Kaigorodov spaces just as $AdS$ spacetimes with a
gravitational wave perturbation.

In this sense, the near horizon limits: M/D-brane $\longrightarrow AdS\times S$ and
M/D-pp-brane $\longrightarrow K \times S$ show a nice
geometric behaviour of the world-volume gravitational waves under the limiting
procedure: indeed the null direction of propagation of the wave on the
brane, $t'$ in (\ref{eq
M2}),  becomes that of the wave on the
Kaigorodov space, ie $t$ in (\ref{kaigo}) or $x^-$ in
(\ref{horo}). In a way these perturbations on the branes can be added
either before or after the near horizon limiting procedure, and yield
a perturbed $AdS$ spacetime. This may be viewed as a decoupling of the
perturbation. \\
On the Penrose limit point of view, any 10 or 11 dimensional
plane-wave spacetime  we
will obtain can trivially be thought of as flat space or a maximally
supersymmetric plane-wave space, with a superposed gravitational wave breaking some of the symmetries. Formally, the analogy between flat membranes and membranes
with a gravitational wave remains both after taking the near-horizon
limit and then taking a Penrose limit which is not trivial. We now
derive the Penrose limits of $K_{n+3}$, and analyse a particular
homogeneous plane-wave we get.

\section{Penrose Limits of $K_{n+3}$}
\label{P limits of K}

\subsection{Construction}
\label{plk}
To classify all the possible Penrose limits of $K_{n+3}$, we could follow the method of ``celestial spheres'' \cite{Blau},
consisting of looking at the orbits of the tangent vectors at a point
under the isotropy subgroup, but this requires finding a group
representation of the Killing vectors. We shall just mention it before
considering a particular limit. Clearly starting from the
metric (\ref{kaigo}) the null geodesics of $K_{n+3}$ which are going to give
non-trivial Penrose limit spaces are those for which $\rho$ varies
with the affine parameter. Otherwise the $\rho$ dependence of the
metric vanishes in the limit and we get flat Minkowski space. In
particular, the limit along the null geodesic of tangent vector
$\partial_t$, ie along $\partial_{x^-}$ the direction of propagation of the
gravitational wave in the horospheres of (\ref{horo}), yields flat space.\\
Because null hypersurfaces play a crucial role in finding coordinates adapted
to taking Penrose limits \cite{Penrose plane-wave}\cite{Penrose}, we adopt the Hamilton-Jacobi
formulation. The following formalism is not explained in
\cite{Penrose}, and it seems that it provides a general way of finding
adapted coordinates of type (\ref{localwave}). Let $S(x,t,\rho,(y^i)) \;$ be such that $g^{\mu
\nu}\partial_{\mu}S\partial_{\nu}S = 0$. Introducing the conserved
``momenta''  $p_x, E, p_i$, we find that  
\begin{equation}
S(x,t,\rho,(y^i))= p_xx+Et+\rho^* +p_iy^i 
\end{equation}
where 
\begin{equation}
\label{geodesic}
\rho^*(\rho, p_x,E,p_i)=\int
\sqrt{e^{-4L\rho}(E^2e^{-(2Ln+4L)\rho}-2Ep_x-p_ip_i)}\, \textrm{d}\rho \equiv \int
f'(\rho)\,\textrm{d}\rho
\end{equation}
Equivalently, the Lagrangian formulation $L=g_{\mu
\nu}\dot{x}^{\mu}\dot{x}^{\nu}=-m^2$ yields
\begin{eqnarray}
\dot{x} &=& Ee^{-4L\rho}, \quad \dot{y}_i=p_ie^{-4L\rho}, \nonumber \\
\dot{t} &=& e^{-4L\rho}(p_x-Ee^{-(4L+2Ln)\rho}),\nonumber \\ 
\dot{\rho}^2 &=& E^2e^{-(8L+2Ln)\rho}-(2Ep_x+p_ip_i)e^{-4L\rho} -m^2
\nonumber \\
\label{lagrange}
\end{eqnarray}
where $m=0$ for null geodesics. We see that $E \neq 0$ is necessary to find null geodesics for which
$\rho$ varies. It is sufficient provided $\rho$ stays small enough.\\
We can easily show here that Kaigorodov spaces admit a
time function, or indeed, because they are time and space orientable,
they are stably causal in the sense of Hawking \cite{Hawking}. The following
argument is adapted from the $4$-dimensional case in \cite{Osinovsky}.We write the
metric (\ref{kaigo}) as 
\begin{displaymath}
ds_{n+3}^2= -(e^{4L\rho+Ln\rho}dt)^2
+(e^{4L\rho+Ln\rho}dt+e^{-Ln\rho}dx)^2+(e^{2L\rho}dy^i)^2 +d\rho^2
\end{displaymath}
and require for timelike or null future-directed geodesics:
\begin{displaymath}
g_{\mu\nu}\dot{x}^{\mu}\dot{x}^{\nu}\leq 0 \quad \textrm{and} \quad
e^{4L\rho+Ln\rho}\dot{t}\geq 0.
\end{displaymath}
If $e^{4L\rho+Ln\rho}\dot{t}$ vanishes at one point, then using
equations (\ref{lagrange}) we see that $\dot{\rho}^2 \geq 0$ implies
$p_x=p_i=E=m=0$ and the
$x^{\mu}$'s must all be constant. Since this is not an acceptable
solution, $\dot{t}>0$ for every timelike or null future-directed
geodesic, and hence $t(\lambda)$ provides a suitable time
function.

We now find the general Penrose limit of (\ref{kaigo}) about the
geodesic of tangent vector $g^{\mu \nu}\partial_{\nu}S =
(\textrm{grad}\,S)^{\mu} \equiv \partial_v $, where $v$ is the affine
parameter along the chosen null geodesic and the future null geodesic
congruence. Note that $\partial_v$ is both orthogonal to the null hypersurfaces $S=u$
and null, hence is tangent to them. It is easier to find an integrable
system of coordinates using covectors or one-forms rather than
coordinate vectors, since the integrability conditions of one forms
are just that they be total derivatives. We want to put (\ref{kaigo}) in the form
\begin{equation}
\label{localwave}
ds^2 = 2du(dv+\frac{1}{2}adu+b_{\alpha}dx^{\alpha})+g_{\alpha
\beta}dx^{\alpha}dx^{\beta}
\end{equation}
with $a$, $b_{\alpha}$ and $g_{\alpha \beta}$ functions of the
coordinates.This is equivalent to looking for coordinates $u$, $v$,
$x^{\alpha}$ such that
\begin{displaymath}
g^{uu}=0, \quad g^{uv}=1, \quad g^{u\alpha}=0
\end{displaymath}
or coordinate-forms such that
\begin{equation}
\label{orthogonality}
\langle du|du \rangle =0, \quad \langle du|dv \rangle =1, \quad
\langle du|dx^{\alpha} \rangle = 0
\end{equation}
A solution to this system is 
\begin{eqnarray}
du&=&p_xdx+Edt+f'(\rho)d\rho+p_idy_i \nonumber \\
\label{ro} dv&=&\frac{d\rho}{f'(\rho)} \\
dz&=&\frac{dx}{E} -\frac{e^{-4L\rho}}{f'(\rho)}d\rho \nonumber \\
dx^i&=&\frac{dy^i}{p_i}-\frac{e^{-4L\rho}}{f'(\rho)}d\rho \nonumber
\end{eqnarray}
The line element then reads:
\begin{eqnarray}
ds^2_{n+3}&=&
2du\big(dv+e^{4L\rho}dz\big)+\big(E^2e^{-2Ln\rho}-2Ep_xe^{4L\rho}\big)dz^2
\nonumber \\
\label{general} & & \quad -2e^{4L\rho}p_idzdx^i+ e^{4L\rho}p_i^2(dx^i)^2
\end{eqnarray}
where $\rho$ is a function of $v$ defined by (\ref{ro}). The Penrose
limit is taken \cite{Blau} by letting $\Omega \longrightarrow 0$ in the following coordinate rescalings: 
\begin{displaymath}
v\rightarrow{v},\quad u
\rightarrow{ \Omega^2 u},\quad  z\rightarrow{\Omega z},\quad
x^i\rightarrow{\Omega x^i}.
\end{displaymath}
The metric is also rescaled by
a factor of $\Omega^{-2}$. Thus the general metric of all non-trivial
Penrose limits of $K_{n+3}$ is 
\begin{eqnarray}
ds^2_{n+3}&=&
2dudv+\big(E^2e^{-2Ln\rho}-2Ep_xe^{4L\rho}\big)dz^2
\nonumber \\
\label{generalpenrose} 
& & \quad -2e^{4L\rho}p_idzdx^i+ e^{4L\rho}p_i^2(dx^i)^2
\end{eqnarray} 
Although this formula breaks down when one $p_i=0$, it is easy to see
that it is equivalent to cancelling the $dzdx^i$ cross term and
keeping the $(dx^i)^2$ term with $p_i=1$. Furthermore, it is clear
that we recover Minkowski space if $\rho(v)$ is constant, but  unlike the 
$AdS$ case, we also get non-trivial limits.  \\
Explicit solutions can be found by integrating and inverting
(\ref{ro}), where $f'(\rho)$ is defined in (\ref{geodesic}). As
explained in \cite{Blau}, Penrose limits taken along null geodesics
related by an isometry  are themselves isometric. Thus given a point
in the initial space, it is sufficient to look at the limits along one 
(rescaled) null vector of each orbit of the ``celestial sphere'' of null vectors under the
isotropy subgroup of the point. As $K_{n+3}$ is homogeneous, one can
choose the origin of the coordinate system in (\ref{kaigo}),
whereupon it is easily seen using the expression of the Killing
vectors in (\ref{killingvectors}) that the isotropy subgroup is generated by the
$L_i$'s and $L_{ij}$'s, and hence is isomorphic to $\mathbb{R}^n \rtimes
\mathcal{SO}(n)$. A simple computation shows that the action of the
$L_i$'s suffices to independently set all the $p_i$'s to zero, but not
$p_x$. One can set $p_x=0$ keeping one $p_i\neq 0$, and the equations
are then related to those in section \ref{winding geodesics
section}. \\
For now, we consider the null geodesic with $p_x=p_i=0$. We get 
\begin{displaymath}
\rho(v)=\frac{1}{L(n+4)}\ln{\big(EL(n+4)v\big)}
\end{displaymath}
and after various rescalings of the coordinates the metric reads: 
\begin{equation}
\label{penrosekaigo} 
ds^2_{n+3}=2dudv+\big(\frac{1}{v}\big)^{\frac{2n}{n+4}}dz^2+v^{\frac{4}{n+4}}dx^idx^i
\end{equation}
Note that this metric has a scaling symmetry:
\begin{equation}
\label{scaling symmetry}
v\mapsto \lambda v, \quad u\mapsto \lambda ^{-1}u, \quad z\mapsto \lambda^{\frac{n}{n+4}}z, \quad x^i\mapsto \lambda^{-\frac{2}{n+4}}x^i.
\end{equation}
The Penrose limit obtained here is valid globally for $v>0$ and $u,
z, x^i \in \mathbb{R}$ : although the
limiting process is defined locally, the coordinate system used to
obtain (\ref{penrosekaigo}) describes the whole initial
spacetime. However the plane-wave metric is well defined for $v<0$, so the
question of coordinate extension through $v=0$ should be raised: the
vanishing of the determinant of the metric at $v=0$ merely signals the presence of
conjugate points.  The coordinate extension 
is usually done by going to Brinkman coordinates. Indeed, for the
maximally supersymmetric solutions, the $\cos v$ coordinate singularity in Rosen
coordinates disappears in Brinkman coordinates \cite{Blau}. \\

\subsection{Geometry and symmetries of the pp-wave}
\label{geometry of the pp-wave}
It is simple to express (\ref{penrosekaigo}) in Brinkman coordinates as the tranverse part of the metric is
diagonal. Consider the following change of coordinates 
\begin{gather}
v=2x^-, \qquad
u=x^+-\frac{1}{2(n+4)x^-}\big(n(y^0)^2-2\sum_{i=1}^{n}{y^i}^2\big),\nonumber
\\
z=(2x^-)^{\frac{n}{n+4}}y^0, \qquad
x^i=(2x^-)^{\frac{-2}{n+4}}y^i \nonumber
\end{gather}
where upon (\ref{penrosekaigo}) becomes
\begin{equation}
\label{brinkmankaigo}
ds^2_{n+3}=2dx^+dx^-+\frac{2n+4}{(n+4)^2}\frac{1}{{x^-}^2}\Big(n(y^0)^2-\sum_{i=1}^{n}{y^i}^2\Big)(dx^-)^2+\sum_{i=0}^{n}{dy^i}^2
\end{equation}
The dependence on $n$ cannot be scaled out. It is easily seen, as
was expected, that
this plane-wave metric satisfies the Einstein vacuum
equations with zero cosmological constant: the $(dx^-)^2$ term,
written as $A_{\mu \nu}y^{\mu}y^{\nu}$, satisfies $\textrm{Tr}(A)=0$. It describes a
gravitational wave propagating in the null direction $x^+$,
distributed along the $y^{\mu}$'s. The null Killing vector $\partial_{x^+}$ is convariantly
constant. We notice that the coordinate singularity at $v=2x^-=0^+$
remains, but although (\ref{brinkmankaigo}) is
ill-defined and the components of the Riemann
tensor diverge as $x^- \to 0$, the square of the Riemann tensor
vanishes. The singularity is a so-called pp-singularity. 
\begin{equation}
\label{Riemann}
R_{-\alpha
-\beta}=\partial_{\alpha}\partial_{\beta}\bigg\{\frac{2n+4}{(n+4)^2}\frac{1}{{x^-}^2}\Big(n(y^0)^2-\sum_{i=1}^{n}{y^i}^2\Big)\bigg\}
\propto \frac{1}{{x^-}^2}
\end{equation}
Actually $x^- \to 0^+$ corresponds to $\rho \to -\infty$ in Kaigorodov
space, so to the pp-singularity initially present in spacetime. In a
sense it remains after taking the Penrose limit. \\
This singularity can be reached in a finite time \cite{Horowitz}, thus the spacetime
is geodesically incomplete. However, the extra scaling symmetry
(\ref{scaling symmetry}), or $x^{+} \mapsto \sigma x^{+}$ and $x^{-} \mapsto
\sigma ^{-1}x^{-} $ in (\ref{brinkmankaigo}), makes this spacetime homogeneous. We now analyse its Lie
algebra of 
symmetries, explain why the space is a Lorentzian homogeneous
space. We then exhibit an In\"on\"u-Wigner contraction
of the Lie algebra  of the initial space into a subalgebra of the
symmetries of the plane-wave spacetime. \\

Applying the general procedure of \cite{Blau} the plane-wave metric
(\ref{penrosekaigo}) admits,
in addition to the $n(n-1)/2$ Killing vectors $e_{ij}$ generating the $\mathcal{SO}(n)$
symmetry algebra of the $x^i$'s,  $2n+3$ Killing vectors spanning a
Heisenberg algebra. Moreover, the scaling symmetry (\ref{scaling
symmetry}) provides an extra Killing vector, $e^-$. In the Rosen
coordinates of (\ref{penrosekaigo}) these vectors read:
\begin{align}
e_+ &= \frac{\partial}{\partial u},\quad e_0=\frac{\partial}{\partial z}, \quad
e_i=\frac{\partial}{\partial x^i},  \nonumber \\
e^*_0 &= z\frac{\partial}{\partial u}
-\Big(\frac{n+4}{3n+4}\Big)v^{\frac{3n+4}{n+4}} \frac{\partial}{\partial z}, \quad 
e^*_i=x^i \frac{\partial}{\partial u}-\Big(\frac{n+4}{n}\Big)v^{\frac{n}{n+4}}
\frac{\partial}{\partial x^i}  \nonumber \\
e_{ij} &=x^i\frac{\partial}{\partial x^j}-x^j\frac{\partial}{\partial
x^i} \nonumber \\
e_- &=v\frac{\partial}{\partial v} -u\frac{\partial}{\partial u}
+\Big(\frac{n}{n+4}\Big)z\frac{\partial}{\partial z}
-\Big(\frac{2}{n+4}\Big)x^i \frac{\partial }{\partial x^i} \nonumber \\
\end{align}
The non trivial commutation relations are $(i,j,k,l=1,\ldots,n)$:
\begin{align}
\label{heisenberg}[e_{\mu}, e^*_{\nu}] &=\delta_{\mu \nu}e_+, \quad \textrm{for} \quad
\mu,\nu =0,1,\ldots,n. \\
 [e_-, e_+] &=e_+, \quad [e_-, e_0]=-\Big(\frac{n}{n+4}\Big)e_0, \quad [e_-,
 e_i]=\Big(\frac{2}{n+4}\Big)e_i, \nonumber \\
\label{outer auto}[e_-, e_o^*] &=\Big(\frac{2n+4}{n+4}\Big)e_0^*, \quad [e_-,
 e_i^*]=\Big(\frac{n+2}{n+4}\Big)e_i^*,  \\
[e_{ij}, e_{kl}] &= -\delta_{ik}e_{jl} + \delta_{jk}e_{il}
 -\delta_{jl}e_{ik} + \delta_{il}e_{jk}, \nonumber \\
\label{vectors} [e_{ij}, e_k] &=\delta_{jk}e_i -\delta_{ik}e_j, \quad
[e_{ij}, e_k^*]=\delta_{jk}e_i^* -\delta_{ik}e_j^*.
\end{align}

Thus $\mathcal{H}=<e^+,e_{\mu}, e_{\nu}^*>$ is a $2n+3$ dimensional
Heisenberg algebra of central element $e^+$, and $\mathcal{SO}(n)$
acts on the $e_i$'s and $e^*_i$'s as on vectors. We notice that
$\mathcal{H}$ and $SO(n)$ generate a Lie algebra
$\mathcal{G}~=~\mathcal{H}(2n+3)\rtimes\mathcal{SO}(n)$ of dimension
$n(n+3)/2 +3$, which is
precisely the same dimension  as $\mathcal{K}$, the symmetry
algebra of $K_{n+3}$.\\
However, there is an extra Killing vector $e^-$, which acts on $\mathcal{G}$
as an outer automorphism. The maximal Lie algebra of symmetries of the plane-wave can be
written as $\widetilde{\mathcal{G}}=\mathcal{H}(2n+3)\rtimes \big(\mathcal{SO}(n)\oplus
\mathbb{R}\big)$, as $e^-$ acts non-trivially only on $\mathcal{H}$. There is
an enhancement of 1 bosonic symmetry in the Penrose limit, while the
fraction of supersymmetry goes from 1/4 to 1/2. \\
In the same way as  homogeneity of the CW-spaces
relies on the extra Killing vector $\partial_{x^-}$, the existence of
the Killing vector $e_+$ or $X=x^+\partial _+ -x^- \partial _-$ implies that the plane-wave spacetime
(\ref{brinkmankaigo}) is (Lorentzian) homogeneous since the other
Killing vectors are clearly transitive on the plane of constant $v$ in
(\ref{penrosekaigo}). Homogeneity
is not hereditary in Penrose limits, and can be lost as we shall see
in the next section. Strictly speaking though, one must
remove the hyperplane $x^-=0$ because it is invariant under the action of
$X$, but it precisely corresponds to the pp-singularity of the
plane-wave and the initial space. Moreover,
$x^->0$, $x^+, y^{\mu} \in \mathbb{R}$ covers the whole initial
Kaigorodov space.

\subsection{In\"on\"u-Wigner contraction}
Although symmetry is enhanced, we can try to relate $\mathcal{K}$ to
$\mathcal{G}$, since they have the same dimension. 
The rescalings of the coordinates by $\Omega$ in the Penrose
Limit suggest an In\"on\"u-Wigner contraction \cite{Inonu} of $\mathcal{K}$ into $\mathcal{G}$. The forthcoming contraction is very
similar to \cite{Hatsuda}
\begin{displaymath}
\mathcal{SO}(3,2)\oplus\mathcal{SO}(8)\longrightarrow{\mathcal{H}(19)\rtimes(\mathcal{SO}(3)\oplus\mathcal{SO}(6)\oplus\mathbb{R})} 
\end{displaymath}
in the Penrose limit $AdS_4\times S^7 \longrightarrow{CW_{max SUSY}}$,
and the other similar types,
apart from the fact that we cannot take the outer automorphism of the
plane-wave algebra since it stems from a symmetry enhancement. \\
$\mathcal{K}$ is       
spanned by the following Killing vectors \cite{Cvetic} expressed in the coordinates of
(\ref{kaigo}):
\begin{align}
K_0 &=  \frac{\partial}{\partial t}, \qquad
K_x=\frac{\partial}{\partial x}, \qquad K_i= \frac{\partial}{\partial
y^i}, \nonumber \\
L_i &= x\frac{\partial}{\partial y^i}-y^i\frac{\partial}{\partial t},
\qquad L_{ij}= y^i\frac{\partial}{\partial
y^j}-y^j\frac{\partial}{\partial y^i}, \nonumber \\
J &= \frac{\partial}{\partial \rho} -(n+4)Lt\frac{\partial}{\partial
t} +nLx\frac{\partial}{\partial x} -2Ly^i\frac{\partial}{\partial y^i}
\nonumber \\ \label{killingvectors}
\end{align}
We define $P_{\pm}=K_0 \pm K_x$ and consider 
\begin{gather}
P_+ \rightarrow{\Omega^2P_+}, \qquad P_- \rightarrow{\Omega P_-}, \qquad
J \rightarrow{\Omega J}, \nonumber \\
K_i \rightarrow{\Omega K_i}, \qquad L_j \rightarrow{\Omega L_j}, \qquad
L_{ij} \rightarrow{\Omega L_{ij}} \nonumber
\end{gather} 
Taking $\Omega \longrightarrow{0}$, the commutation relations of
$\mathcal{K}$ (see \cite{Cvetic}) yield
those of $\mathcal{G}$, where we further make the following association
\begin{gather}
e_+=P_+, \qquad e_0=P_-, \qquad e^*_0=-\frac{J}{(n+2)L}, \nonumber \\
 e_i=-2K_i,\qquad e^*_i=L_i, \qquad e_{ij}=L_{ij}.
\end{gather}
Thus we get the In\"on\"u-Wigner contraction
$\mathcal{K}\longrightarrow \mathcal{H}(2n+3)\rtimes
 \mathcal{SO}(n)$. However one could perfectly interchange $P_+$ and $P_-$ and get $P_-$ as
the central element of the contracted algebra: it is not clear
whether $\mathcal{K}$ really ``undergoes'' a contraction in the
Penrose limit, as the latter is not uniquely defined. The similar
feature arises in the maximally supersymmetric cases. Nevertheless,
one likes to think of Penrose limits as yielding approximations of
spacetimes, and the trivialization of any algebra of symmetries into a
Heisenberg type algebra certainly ``happens''. On the dynamical point
of view, a precise understanding of what happens group theoretically
should help to explain how the degrees of motion of a particle or a
string decouple to yield independent
harmonic oscillators in the plane-wave limit. We  now
consider possible limits of the whole supergravity or IIB spacetimes,
topological products of Kaigorodov spaces and spheres.

\section{Penrose limits of $K_{n+3}\times S^d\;$spaces}
\label{general penrose limits}
We want to find a coordinate system which singles out a particular
null geodesic congruence of the following metrics, for $n+3+d=10$ or 11:
\begin{equation}
ds^2=e^{-2Ln\rho}dx^2+e^{4L\rho}(2dxdt+dy^idy^i)+d\rho^2+d\psi^2+(Ln)^{-2}\sin^2\!(Ln\psi)
d\Omega^2_{d-1}
\end{equation}
Again we adopt the Hamilton-Jacobi
formulation, with $S(x,t,\rho,(y^i),\psi)$ a function of the coordinates
satisfying  $g^{\mu
\nu}\partial_{\mu}S\partial_{\nu}S = 0$. Introducing $l$, angular
momentum about $\psi$, in addition to the same conserved quantities as
before,  we find 
\begin{equation}
S(x,t,\rho,(y^i),\psi)= p_xx+Et+ p_iy^i+\rho^* +l\psi 
\end{equation}
where now
\begin{equation}
\label{geod}
\rho^*(\rho, p_x,E,p_i,l)=\int
\sqrt{e^{-4L\rho}(E^2e^{-(2Ln+4L)\rho}-2Ep_x-(p_i)^2 )-l^2}\, \textrm{d}\rho \equiv \int
f'(\rho)\,\textrm{d}\rho
\end{equation}

\subsection{Non-winding geodesics}
We will first consider the case where the chosen null geodesic of
tangent vector $g^{\mu \nu}\partial_{\nu}S =
(\textrm{grad}\,S)^{\mu} \equiv \partial_v $ does not wind around the
sphere, ie $l= 0$, $E\neq 0$. In this case, the spherical part of the metric
yields flat space, and the limits of $K_{n+3}$ where found in the
previous section. In the particular case
$p_x=p_i=0$, using (\ref{brinkmankaigo}), the limit metric is simply: 
\begin{equation}
\label{nonwinding}
ds^2_{n+3+d}=2dx^+dx^-+\frac{2n+4}{(n+4)^2}\frac{1}{{x^-}^2}\big(n(y^0)^2-\sum_{i=1}^{n}{y^i}^2\big)(dx^-)^2+\sum_{i=0}^{n}{dy^i}^2+ds^2_{\mathbf{R}^d}
\end{equation}
The field strength G\"uven \cite{Gueven} limit
$ F^*_{p+1}=\lim_{\Omega \to {0}}\Omega^{-p}F_{p+1}$ yields
$F_{4}=0$ for $K_4\times S^7$, whatever $p_x$,
$p_1$, provided that  $E\neq 0$ and $l=0$. Indeed 
\begin{displaymath}
F_4 \propto{e^{6L\rho}f'({\rho})du\wedge dz\wedge dy^1\wedge dv}
\end{displaymath}
so $F_4$ goes as $\Omega$ in the limit and vanishes. The same is true
for the field strengths $\ast[F_4]$ and $F_5$.
This merely reflects the fact that Kaigorodov
spaces are solutions of Einstein pure gravity. Indeed, as a consequence, their Penrose limits,
and hence the Penrose limits of $K \times S$ spaces along \emph{non winding}
geodesics, satisfy Einstein pure gravity too. Therefore, all the plane-wave
spaces
of type (\ref{nonwinding}) for $n+3+d=11$ or 10  are trivial solutions of
supergravity (with vanishing field strengths). They describe a gravitational wave propagating in $n+3$
dimensions. However, they only arise as limits of brane-like solutions
for $n=1$, 2 or 4. \\
As all generic pp-waves, the spaces (\ref{nonwinding}) preserve $1/2$ of the
supersymmetries, merely the constant spinors $\epsilon$ satisfying
$\Gamma_+\epsilon=0$ in the light-cone vielbein formalism (see
\cite{Hull}\cite{Pope} for example). Since the field strengths vanish, there are no
non-constant solutions and these are the only solutions. Comparing
with the $AdS \times S$ Penrose limits, we here have non-trivial gravitational
waves even in the non-winding case. They are essentially $n+3$
dimensional, and thus can be viewed as consequences of the waves perturbing the
initial branes and their near horizon geometries. Formally one can write
(\ref{nonwinding}) as a Minkowski metric (Penrose limit of the $AdS\times S$
along non-winding geodesics) ``perturbed'' by an $n+3$ dimensional wave. \\
In fact these plane-wave spacetimes, as their $n+3$ dimenesional
analogues (\ref{brinkmankaigo}), are Lorentzian homogeneous spaces. In
10 dimensions, with a constant dilaton field, they can provide a time
dependent background on which string theory is exact and solvable
\cite{Loughlin}. However, the term $n(y^0)^2$ will give rise to a
negative mass harmonic oscillator. Fixing $y^0=0$ corresponds to the
background studied in \cite{Russo}.     

\subsection{Winding geodesics}
\label{winding geodesics section}
Whereas in the previous case the plane-wave obtained propagated in
$n+3$ dimensions only, we might expect as in the $AdS\times S$ cases
\cite{Blau} to find an essentially
11(or 10)-dimensional plane-wave by taking a null geodesic which winds
around an equator of the sphere. Anagously to section (\ref{plk}) a possible coordinate transformation
for $E\neq 0$, $l\neq0$, $p_x=p_i=0$ is given by :
\begin{eqnarray*}
du&=& Edt + f'(\rho)d\rho +ld\psi \\
dv&=& \frac{d\rho}{f'(\rho)} \\
dw&=& - \frac{d\rho}{f'(\rho)}+ \frac{d\psi}{l} \\
dz&=& \frac{dx}{E}-\frac{e^{-4L\rho}}{f'(\rho)}d\rho \\
dy^i&=&dy^i
\end{eqnarray*}
This is only valid for $\rho \leq \frac{1}{2Ln+8L}\ln {(\frac{E}{l})^2}\equiv \rho_0$, because
greater values of $\rho$ are not reached by the chosen geodesic
congruence, as can be seen by analysing (\ref{geod}). Again the
Hamilton-Jacobi formalism to find null hypersurfaces and therefore
possible null geodesic congruences, together with  the orthogonality conditions on the
coordinate-forms (\ref{orthogonality}), yield this non-trivial (integrable) change of
coordinates. The metric then reads 
\begin{eqnarray}
ds^2_{n+3+d}&=&2du(dv+e^{4L\rho(v)}dz)+E^2e^{-2Ln\rho(v)}dz^2+e^{4L\rho(v)}(-2l^2dzdw+dy^idy^i)\nonumber
\\
\label{nearlythere}
 & & +l^2dw^2+\sin^2\!(Lnl(w+v))d\Omega^2_{d-1}, \quad \textrm{where} \\
\label{rofinal} 
\rho(v) &=& \frac{1}{2Ln+8L}\ln{\cos^2\!\big(l(4L+Ln)v\big)}+\rho_0
\end{eqnarray}
To find $\rho(v)$ one must set  
$0\leq-l(4L+Ln)v\leq \pi/2$ during inetegration, however
(\ref{rofinal}) stays valid for all $v\in \mathbb{R}$ if we accept
$\rho$ periodic in $v$ on the range $(\,-\infty, \rho_0 \,]$, and the
remaining coordinate transforms still make sense. Note that (\ref{nearlythere}) does not cover the whole of $K_{n+3} \times S^d$,
but at any given point (so any $\rho_0$), we can pick $E$ and $l$ so
 as to cover a neighbourhood of that point with null geodesic
congruence coordinates which break down when $\cos(-l(4L+Ln)v)=0$. The
coordinates of (\ref{nearlythere}) cannot be
used to discuss global properties of $K_{n+3}\times S^d$ spacetimes.

As usual we define the following rescalings of the coordinates, the
metric and the field strength:
\begin{align}
v &\rightarrow{v}, \quad u\rightarrow{\Omega^2u}, \quad
(z,w,y^i)\rightarrow{(\Omega z, \Omega w, \Omega y^i)}, \nonumber \\
g &\rightarrow{\Omega^{-2} g}, \quad F_{p+1}\rightarrow{\Omega^{-p} F_{p+1}},
\nonumber 
\end{align}
The Penrose-G\"uven limit is obtained by taking
$\Omega\longrightarrow{0}$.  Using (\ref{rofinal}) and
$f'(\rho)=l\tan(-l(4L+Ln)v)$, (take $l\geq 0$) this yields: 
\begin{eqnarray}
ds^2 &=&
2dudv+E^2\big(\frac{E^2}{l^2}\cos^2\!(l(Ln+4L)v)\big)^{-\frac{n}{4+n}}dz^2
+l^2dw^2+\sin^2\!\big(Lnlv)ds^2_{\mathbf{R}^{d-1}} \nonumber \\
 & & +\big(\frac{E^2}{l^2}\cos^2\!(l(Ln+4L)v)\big)^{\frac{2}{4+n}}(-2l^2dzdw
+dy^idy^i)  \\
F_4 &=& 6Ll^2f'(\rho)e^{6L\rho}dv\wedge dw\wedge dz\wedge dy^1
\nonumber \\
    &=& 6Ll^3\tan(-l(4L+Ln)v)\big(\frac{E^2}{l^2}\cos^2\!(-l(4L+Ln)v)\big)^{\frac{3}{n+4}}dv\wedge
dw\wedge dz\wedge dy^1 \nonumber 
\end{eqnarray}
After rescaling the coordinates appropriately by powers of $E$ and
$l$, we get the following plane-wave spacetime, where the flux term $F_4$ is
only relevant in the $n=1$ case (limit of $K_3\times S^7)$: 
\begin{eqnarray}
ds^2 &=& 2dudv+\big(\cos v\big)^{-\frac{2n}{4+n}}dz^2+\big(\cos
v\big)^{\frac{4}{4+n}}(2dzdw
+dy^idy^i) \nonumber \\
\label{finalrosen}
 & & +dw^2+\sin^2\!\big((n/(n+4))v\big)ds^2_{\mathbf{R}^{d-1}}\\
F_4 &=&\frac{6}{n+4}\tan v\big(\cos v \big)^{\frac{6}{n+4}}dv\wedge
dw\wedge dz\wedge dy^1 
\end{eqnarray}
These Rosen coordinates are valid for $0\leq v\leq \pi/2$. Similarly we
obtain from (\ref{fluxes}) the limits of $\ast[F_4]$ and $F_5$ for the
 $n=4$ and $2$ cases respectively:
\begin{eqnarray}
\ast[F_4] &=& \frac{12}{n+4}\tan v\big(\cos
v\big)^{\frac{12}{n+4}}dv\wedge dw\wedge dz\wedge d^4y^i  \quad (n=4)
\\
F_5 &=& \frac{8}{n+4}\tan v\big(\cos
v\big)^{\frac{8}{n+4}}dv\wedge dw\wedge dz\wedge d^2y^i \nonumber \\
  & & +\ast[\frac{8}{n+4}\tan v\big(\cos
v\big)^{\frac{8}{n+4}}dv\wedge dw\wedge dz\wedge d^2y^i] \quad (n=2)
\end{eqnarray}
Note that the metrics and the field strengths obtained are independent of $E$,
$l$, and also the Kaigorodov parameter $L$. Only the region of the initial space covered by
$0\leq v\leq \pi/2$ depends on them. The limit spaces do not
depend on the charges of the initial pp-branes. 

One can easily find the Killing vectors of the metric
(\ref{finalrosen}) which generate a Heisenberg algebra of
dimension $2(n+d+1)+1$, together with those which span the $\mathcal{SO}(n)$ and $\mathcal{SO}(d-1)$ algebras
respectively (see \cite{Blau} for example). Counting these symmetries,
and since $K_{n+3} \times S^d$
has $n(n+3)/2 +3 +(d+1)d/2$ Killing vectors, there is an enhancement
of at least 1 bosonic symmetry in the Penrose limit, and in fact no more. This is true in
the three cases. There is no obvious additional scaling symmetry as we had in
(\ref{penrosekaigo}). As suggested in
\cite{Loughlin}, it is often easier to see that in Brinkman
coordinates.

\subsection{Brinkman coordinates}
The change of coordinates between Rosen and Brinkman coordinates has
been known for a long time. It is reviewed in \cite{Blau}
\cite{Loughlin} for example.  We relabel the
coordinates of (\ref{finalrosen}) by $z= x_0$, $w= x_1$,
$y_i= x_{i+1}$ for $i=1,\ldots,n$, and also call $x_i$, for $i=n+2, \ldots,
n+d$,  the coordinates of $\mathbb{R}^{d-1}$. From now on latin indices
range from 0 to $n+d$.  The spatial part of
(\ref{finalrosen}), denoted $C_{ij}(v)dx^idx^j$,  is not diagonal, and cannot be
made so by staying in Rosen type coordinates. Thus the change of
coordinates involves a particular inverse vielbein field ${Q^i}_j$ of
$C_{ij}$  which non-trivially rotates the $z=x_0$ and $w=x_1$
coordinates. Formally, the solution can be written:
\begin{align}
&v=x^-, \quad y^i= {Q^{i}}_j z^j, \nonumber \\
\label{tobrinkman}&u=x^+ -\frac{1}{2}C_{ij}{\dot{Q}^i}_k{Q^j}_l z^kz^l,
\end{align}
with $Q$ satisfying $Q^TCQ=\mathbb{I}$ and $\dot{Q}^TCQ=Q^TC\dot{Q}$,
where the dot denotes differenciation with respect to $x^+$. This
yields the plane-wave metric:
\begin{align}
ds^2 =& 2dx^+dx^- +A_{ij}(x^-)z^iz^j(dx^-)^2 +dz^idz^i \nonumber \\ 
\label{A's formula} & \textrm{where} \quad
A_{kl}=-(\dot{C_{ij}\dot{Q^i_k}}){Q^j}_l 
\end{align}
A possible solution is given by $Q(v)$ whose only non-vanishing
components are:
\begin{gather*}
{Q^0}_0(v)=(\tan v)^{-1}( \cos v)^{-\frac{4}{n+4}} \cos
\Big(\frac{2v}{n+4}\Big), \\
{Q^0}_1(v)= (\tan v)^{-1}( \cos v)^{-\frac{4}{n+4}} \sin
\Big(\frac{2v}{n+4}\Big) \\
{Q^1}_0(v)= -(\tan v)^{-1} \cos
\Big(\frac{2v}{n+4}\Big) -\sin \Big(\frac{2v}{n+4}\Big), \\ 
{Q^1}_1(v)= -(\tan v)^{-1} \sin
\Big(\frac{2v}{n+4}\Big) +\cos \Big(\frac{2v}{n+4}\Big) \\
{Q^i}_i(v)= (\cos v)^{-\frac{2}{4+n}} \quad \textrm{for} \quad
i=2,\ldots, n+1 \\
{Q^i}_i(v)= \sin^{-1} \Big(\frac{nv}{n+4}\Big) \quad \textrm{for} \quad
i=n+2, \ldots, n+d.
\end{gather*}
The expression for $A_{ij}$ given by (\ref{A's formula}) turns out
complicated (and non-diagonal) for $i,j\in \{0,1\}$. For the diagonal
terms $(i,j>1)$, $A_{ij}(x^-)=
\frac{\ddot{\sqrt{C_{ii}}}}{\sqrt{C_{ii}}}\delta_{ij}$ reads:
\begin{align*}
A_{ii} &=-\Big(\frac{2}{n+4}\Big)^2\big(1+\frac{n+2}{2}\cos^{-2}x^-\big)
\quad \textrm{for} \quad i=2,\ldots, n+1 \\
A_{ii}&= -\Big(\frac{n}{n+4}\Big)^2 \quad \textrm{for} \quad i=n+2,
\dots, n+d.
\end{align*}
Then letting $a,b\in \{0,1\}$, and defining $\widetilde{A}_{ab}(x^-)=A_{ab}(x^-)
+(\frac{2}{n+4})^2 \delta_{ab}$, the metric reads:
\begin{eqnarray}
ds^2 &=& 2dx^+dx^- -\Big(\frac{n}{n+4}\Big)^2 \Big\{
\Big(\frac{2}{n}\Big)^2 \sum_{i=0}^{n+1}(z^i)^2
+\sum_{i=n+2}^{n+1+d}(z^i)^2 \Big\}(dx^-)^2 + dz^idz^i \nonumber \\
\label{we're there} & & \quad + \Big\{ \widetilde{A}_{ab}(x^-)z^az^b -\frac{2n+4}{(n+4)^2}\cos^{-2}x^-
\sum_{i=2}^{n+1}(z^i)^2\Big\}(dx^-)^2.  
\end{eqnarray}
Introducing $\widetilde{A}_{ab}$ seems unecessary. However, the
\emph{first line}  of (\ref{we're there}) describes the maximally
supersymmetric plane-waves (when supported by appropriate field strengths
of course). Indeed, for $n=1$ or $n=4$ in 11 dimensions, ($(2/n)^2=4$ or
$1/4$), it 
corresponds to the Kowalski-Glikman solution \cite{Kowalski} (also
described in \cite{O'Farrill}), while for
$n=2$ in 10 dimensions,  it is the metric of the BFHP maximally
supersymmetric type IIB plane-wave \cite{BFHP}. Although we could say of any plane-wave that it consists of the sum of a
maximally supersymmetric Cahen-Wallach space and another plane-wave,
the argument seems illuminating here in the context of Penrose limits
of particular brane solutions. Indeed, the Penrose limits of the $AdS
\times S$ spaces along geodesics which wind round the sphere
\cite{Blau} yield
the maximally supersymmetric plane-waves. Here, starting from  a $K_{n+3}
\times S^d$ geometry, or an $AdS_{n+3} \times S^d$ spacetime perturbed with an
$(n+3)$-dimensional gravitational wave, we
exhibit a Penrose limit space which can quite naturally be interpreted
as a maximally supersymmetric plane-wave together with an additional
gravitational wave.  
When solving the Kiling equations \cite{Loughlin}, we see that
(\ref{we're there}) does not admit a Killing vector with non-vanishing
$\partial_{x^-}$ component. Indeed, the diagonal terms  $A_{ii}$ for
$i=2, \ldots, n+1$ are neither constants nor proportional to
$(1/x^-)^2$. Hence the plane-wave we obtain is not
homogeneous. Moreover, (\ref{we're there}) tells us that the plane-wave limit has a
pp-singularity at $v=\pi/2$. As in the non-winding case, this
singularity stems from the pp-singularity of the Kaigorodov space,
which itself corresponds to the divergence of the tidal forces as one
approaches the pp-branes of the initial geometries.       

\section{Conclusion}
The plane-wave spacetimes obtained in this paper should really be
thought of as arising from Penrose limits of Kaigorodov spaces,
themselves near horizon limits of M2, M5 or D3-pp-branes. In this
sense they are dynamically relevant. Moreover $K_{n+3}\times S^d$
spaces provide themselves  interesting solutions of supergravity which
are analogous to the $AdS \times S$ configurations. In
\cite{Cvetic} and \cite{Chamblin} evidence was given that gravity in
the Kaigorodov space is dual to a CFT in the infinite momentum frame
with constant (null) momentum density, and that therefore one can
consider the Kaigorodov space as an infinitely boosted version of
$AdS$. When taking Penrose limits, it
seems difficult to
keep track of this fact. However, when we think of $K$ as an
$AdS$ space perturbed by a gravitational wave, the picture seems clearer. The plane-waves
(\ref{nonwinding}) and (\ref{we're there}) can be naturally
interpreted as the corresponding Penrose limits of $AdS\times S$
spaces along non-winding and winding null geodesics, perturbed by a
gravitational wave. This decoupling, similar to the one
occuring when taking the near horizon geometries of pp-branes, needs
to be investigated further. 

Our analysis of various Penrose limits certainly illustrates the variety of possible
symmetry enhancements. The example of the Kaigorodov space limit is
worth noting: an enhancement of one bosonic symmetry, but nevertheless an
In\"on\"u-Wigner contraction of the initial algebra of symmetries to a
subalgebra of the plane-wave symmetries. Note that since all
plane-wave Lie algebras have the same structure (up to a possible 
extra  outermorphism), the contraction of the algebra of $K_{n+3}$ into
$\mathcal{H}(2n+3)\rtimes \mathcal{SO}(n)$ can be shown to occur
systematically in all cases. The loss of homogeneity when
taking the limit of $K\times S$ spaces along winding geodesics might
seem striking, but is linked to the fact that the limiting process is
essentially local. In addition,  although the number of Killing vectors is conserved or
increases \cite{Blau}, their action often becomes redundant. For
example, in the typical $2d+1$-dimensional plane-wave Heisenberg
algebra, possibly enlarged by  a semi-direct product of rotations, only $d+1$ symmetries yield
 motions  of the spacetime in independent dimensions.  The
question of homogeneity becomes important when one wants to solve
string theories in plane-wave backgrounds in view of possibly
relating them  to a certain CFTs, since the extra conserved quantity
simplifies the equations \cite{Loughlin}. The precise understanding of
what the algebra (and super-algebra) of symmetries incurs in the limit, should help to
 understand how the degrees of motion of strings and particles moving
in a given background decouple to yield, in the bosonic part for
example, independent harmonic oscillators, and then better relate their
simple  dynamics to that of the initial space.

\section*{Acknowledgements}
I would like to thank Gary Gibbons for motivating discussions and
insightful comments, and also the EPSRC, the DAMTP, and the Cambridge
European Trust for financial support. I am also
grateful to the referees for some useful comments.

\appendix
\section{Isometric embedding of the Kaigorodov space}
\label{embedding of kaigo}
It seems natural to look for an embedding of $K_{d+1}$ which
resembles that of $AdS_{d+1}$ in Minkowski space of signature
$(2,d)$. Consider the $(2,d)$-signature space of metric ($i\in
\{1,\ldots ,d-2\}$): 
\begin{equation}
\label{embedding}
ds^2=dUdV+dX^+dX^- +dX^idX^i+R^d\Big(\frac{{dX^+}^2}{U^d}+\frac{{X^+}^2}{U^{d+2}}du^2-2\frac{X^+}{U^{d+1}}dUdX^+\Big)
\end{equation}
One can isometrically embed the hyperboloid-like hypersurface defined
by:
\begin{equation}
\label{hyper}
UV+X^+X^-+(X^i)^2=-R^2
\end{equation}
Analogously to the $AdS$ case we define horospherical coordinates on this
hypersurface:
\begin{displaymath}
z=\frac{R}{U}, \quad x^{\pm}=X^{\pm}\frac{z}{R}, \quad x^i=X^i\frac{z}{R}
\end{displaymath}
and replace $V(U,X^{\pm},X^i)$ with (\ref{hyper}). The induced metric
on the hypersurface is then that of $K_{d+1}$ in horospherical-like
coordinates, for $z>0$ :
\begin{equation}
\label{horo2}
ds^2_{d+1}=\frac{R^2}{z^2}\big(2dx^+dx^- +z^{d}(dx^+)^2 +\sum_{i=1}^{d-2}{dx^i}^2+dz^2\big)
\end{equation}
One might think that a simpler embedding space could be found by
changing the right-hand-side of (\ref{hyper}) by a function
$f(U,X^+)$, but in  fact it only amounts to a change of coordinates in
the embedding space.
The embedding space satisfies the Einstein vacuum field equations, is
Ricci flat, and
its Riemann and Weyl tensors only have the following independent
non-vanishing components:
\begin{displaymath}
R_{UX^+UX^+}=C_{UX^+UX^+} \propto{\frac{1}{U^{d+2}}}
\end{displaymath}
Therefore this space is not a symmetric space (the Riemann tensor is not covariantly constant). As a side remark, solving Killing's equations of
(\ref{embedding}) should not be difficult, and finding 
the Killing vectors which survive on the hypersurface might give a deeper
insight into the Lie group of motions of the generalized Kaigorodov
spaces, as an intersection of the anti-de-Sitter group and the group
of motions of (\ref{embedding}).

\end{document}